\documentclass[bibyear]{aa} 

\usepackage{graphicx}
\usepackage{txfonts}
\begin{document}
   \title{Vertical velocities from proper motions of red clump giants}

   \subtitle{}

   \author{
         M. L\'opez-Corredoira\inst{1,2}, H. Abedi\inst{3}, F. Garz\'on\inst{1,2},
F. Figueras\inst{3} }
\institute{
$^1$ Instituto de Astrof\'\i sica de Canarias,
E-38205 La Laguna, Tenerife, Spain\\
$^2$ Departamento de Astrof\'\i sica, Universidad de La Laguna,
E-38206 La Laguna, Tenerife, Spain\\
$^3$ Dept. d’Astronomia i Meteorologia, Institut de Ci\`encies del Cosmos, 
Universitat de Barcelona, IEEC,
Mart\'\i \ i Franqu\'es 1, E08028 Barcelona, Spain}

\offprints{martinlc@iac.es}
\titlerunning{Vertical velocities / Red Clumps}
\authorrunning{L\'opez-Corredoira et al.}

   \date{Received xxxx; accepted xxxx}

 
  \abstract
    {}  
  {We derive the vertical velocities of disk stars in the range of Galactocentric radii of $R=5-16$ kpc
within 2 kpc in height from the Galactic plane. This kinematic information is connected to dynamical
aspects in the formation and evolution of the Milky Way, such as the passage of satellites and
vertical resonance and determines whether the warp is a long-lived or a transient feature.}
  {We used the PPMXL survey. To improve the accuracy of the proper motions, the systematic shifts from zero were calculated by using the average proper motions of quasars
in this PPMXL survey, and we applied the corresponding correction to the proper motions of the whole survey, which reduces the systematic error. From the color-magnitude diagram $K$ versus $(J-K)$ we selected the standard candles corresponding to red clump giants and used the information of their proper motions to build a map of the vertical motions of our Galaxy. 
We derived the kinematics of the warp both analytically and through a particle simulation to fit these data. Complementarily, we also carried out the same analysis with red clump giants spectroscopically selected with APOGEE data, and we predict
the improvements in accuracy that will be reached with future Gaia data.}
  {A simple model of warp with the height of the disk
$z_w(R,\phi )=\gamma (R-R_\odot )\sin (\phi -\phi _w)$ fits the vertical motions if
$\dot{\gamma }/\gamma =-34\pm 17\,{\rm Gyr}^{-1}$; the
contribution to $\dot {\gamma }$ comes from the southern warp and is negligible in the north.
If we assume this 2$\sigma $ detection to be real, the period of this oscillation is shorter than 0.43 Gyr at 68.3\% C.L. and shorter than 4.64 Gyr at 95.4\% C.L., which excludes long-lived features. Our particle simulation also indicates a probable abrupt decrease of the warp amplitude in a time of about one hundred Myr.}
  {The vertical motion in the warp apparently indicates that the main S-shaped structure of the warp is a long-lived feature, whereas the perturbation that produces an irregularity in the southern part is most likely a transient phenomenon. With the use of the Gaia end-of-mission products together with spectroscopically classified red clump giants, the precision in vertical motions can be increased by an order of magnitude at least.}

   \keywords{Galaxy: kinematics and dynamics --- Galaxy: disk}

   \maketitle
%

\section{Introduction}

Average vertical motions (i.e., motions perpendicular to the Galactic plane) of disk stars may be produced by different dynamical causes, such as the passage of satellites (Widrow et al. 2012) or vertical resonance (Griv et al. 2002) or the warp. 
In case of the warp, we have to distinguish between the vertical component of
the velocity caused by the inclination of the orbit of the stars with respect to the Galactic plane, and the possible evolution of inclination of the warp angle. Whether there is such an evolution
or not distinguishes the scenarios in which the warp is a long-lived or a transient feature.
Theories that predict a long-lived warp see a steady torque as the reason (for instance, produced by a steady accretion of matter onto the disk, L\'opez-Corredoira 2002a). Transient warps may be deduced when the torque is variable (for instance, when the torque is produced by the misalignment of halo and disk and when the realignment is produced in $\lesssim 1$ Gyr; Jiang \& Binney 1999).
These ideas can be testd by measuring of vertical motions for stars at large
Galactocentric radii; this is poorly known field of research as yet.

Some analyses of the warp kinematics were produced 
by Miyamoto et al. (1993), Miyamoto \& Zhu (1998), Smart et al. (1998), 
Drimmel et al. (2000), Makarov \& Murphy (2007), and Bobylev (2010), but all analyses were made 
within short distances from the Sun and those that used
data from Hipparcos were subject to some systematic errors in the parallaxes and proper motions
(Drimmel et al. 2000; Momany et al. 2006). Larger distances were explored by Bobylev (2013) using 
Cepheids as standard candles, but the number of stars was small and his analysis may have different
interpretations (see discussion in Sect. \ref{.warp}). 

Here we add further information and constraints to our knowledge on vertical motions.
The tool we use for our analysis of kinematics is a catalog of proper motions.
Bond et al. (2010) and Widrow et al. (2012) have also made an extended analysis of the stellar 
kinematics from Sloan Digital Sky Survey (SDSS) data, but these analyses do not cover the Galactic plane at $|b|<20^\circ $, 
therefore it cannot be used to explore large Galactocentric distances. 
Here we concentrate on the region $|b|<20^\circ $.

We use the red clump giant (RCG) population, which is a very appropriate standard candle (e.g., Castellani et al. 1992) with relatively low errors on its distance estimate. It can be observed up to distances of $\approx 8$ kpc, which allows one to reach $R\lesssim 16$ kpc (L\'opez-Corredoira et al. 2002b). Williams et al. (2013) have used RCGs to derive this kinematic information, but only within the range  $6<R($kpc$)<10$, $|z|<2$ kpc, which does not reach the farthest regions of the stellar disk we aim to explore here: deriving the vertical motions from proper motions of RCGs in the range $4<R($kpc$)<16$ for $|z|<2$ kpc. 

The structure of this paper is as follows: in Sect. \ref{.data} the data source is described;
the method of finding the RCG stars is detailed in Sect. \ref{.redclump}; 
the vertical velocity from proper motions is calculated in Sect. \ref{.proper} for PPMXL data, 
which leads to the results and their dynamical consequences in Sect. \ref{.warp} and \ref{.waves}.
Regrettably, the measurements of the vertical
motions are only moderately accurate, so further research will be needed to establish a more precise constraint. In Sect. \ref{.improve} we indicate ways of improving our results with 
future surveys. A summary and conclusions are presented in the last section.
We assume for the Sun a Galactocentric distance of 8 kpc, 
a circular/azimuthal velocity of 250 km/s, and a flat rotation speed for the disk of our Galaxy of 238 km/s.

\section{Data from the star catalog PPMXL: subsample of 2MASS}
\label{.data}

The star catalog PPMXL (Roeser et al. 2010) lists positions and proper motions of
about 900 million objects and is complete for the whole sky down to
magnitude $V\approx 20$. It is the result of the re-reduction of the catalog
of astrometry, visible photometry and proper motions of the USNO-B1 catalog  
cross--correlated with the astrometry and near-infrared photometry of the 2MASS point-source
catalog, and re-calculating the proper motions in the absolute reference frame of
International Celestial Reference Frame (ICRS), with respect to the barycenter
of the solar system. The typical statistical errors of these proper motion
is 4-10 mas/yr, while the systematic errors are on average 1-2 mas/yr.
From the whole PPMXL, we first selected the subsample of 2MASS sources with 
$m_K\le 14$ and available $J$ photometry. This yielded 
a total of 126\,636\,484 objects with proper motions, that is, 
an average of 3\,100 sources deg$^{-2}$.

There is another star catalog that might be appropriate for this analysis: 
UCAC4 (Zacharias et al. 2013) lists positions and proper motions of
about 105 million objects, aiming to be complete for the whole sky down to
magnitude $R\approx 16$. Like PPMXL, it also gives proper motion in ICRS reference frame.
The UCAC is the first modern high-density, full-sky star catalog that is not based on photographic images of the sky for most stars, but on recent CCD observations, with higher accuracy in the differential positions. However, given the shorter period of observation of stars, the accuracy of the proper motions
is similar to PPMXL: the typical statistical errors of these proper motion
is 1-10 mas/yr, while the systematic errors are on average 1-4 mas/yr. The main disadvantage of this catalog
and the reason why we here selected PPMXL as the survey for our analysis is that the limiting
magnitude of $R\approx 16$ does not allow estimating the systematic errors, as we do with
PPMXL data in Sect. \ref{.corrsyst} because there are very few quasars up to that limiting magnitude. Nonetheless, we are preparing a paper that uses UCAC4 data to derive some information
on vertical motions for whihch we apply a different method of systematic errors correction: Abedi et al. (in preparation).

\section{Selection of red clump giants}
\label{.redclump}

We selected the red clump giants (RCGs) in $K$ versus $J-K$ color-magnitude
diagrams. Near-infrared color magnitude diagrams are very suitable for separating the
RCGs because they clearly separate the dwarf population up to magnitude $m_K\lesssim 13$
(L\'opez-Corredoira et al. 2002b). Moreover, the absolute magnitudes vary only little in the
near-infrared with metallicity or age (Castellani et al. 1992; 
Salaris \& Girardi 2002; Pietrzy\'nski et al. 2003). 
This narrow luminosity function distribution (Castellani et al. 1992) makes them very appropriate standard candles that trace the old stellar population of the Galactic structure.

\begin{figure}
\centering
\vspace{1cm}
\includegraphics[width=8cm]{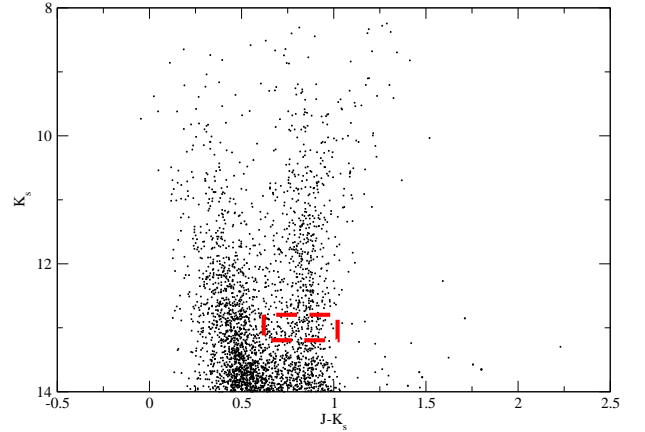}
\caption{Example of applying the RCG selection from the PPMXL survey in a $K$ vs. $J-K$ color-magnitude diagram.
This case is for the direction $\ell=180^\circ $, $b=7^\circ $, at one particular K-magnitude 
$m_{K,0}=13.0$, giving the highest counts
within the RCG region at $(J-K)_0=0.81$. The dashed rectangle represents the area where the RCGs are selected.}
\vspace{1cm}
\label{Fig:CMl180b7}
\end{figure}

Here we used the same selection method for RCGs as in L\'opez-Corredoira et al. (2002b).
For each magnitude $m_{K,0}$ we determined the $(J-K)_0$ that yields the highest count density along this horizontal cut of the color-magnitude diagram for the RCG, and we considered all the stars within the rectangle with points $[(J-K), m_K]$ such that 
\begin{equation}
m_{K,0}-\Delta m_K \le m_K \le m_{K,0}+\Delta m_K
,\end{equation}\begin{equation}
Max[(J-K)_{{\rm red\ clump}},(J-K)_0-\Delta (J-K)]\le (J-K)
\end{equation}\[
\,\,\,\,\,\le (J-K)_0+\Delta (J-K)
.\] 
We take $\Delta m_K=\Delta (J-K)=$0.2 mag. An example is given in Fig. \ref{Fig:CMl180b7}. 
Here, we ran the same application for the whole area of the sky with $|b|\le 20^\circ $ in bins of $\Delta \ell =1^\circ$, $\Delta b=1^\circ $ and $9.8<m_{K,0}\le 13.0$ in bins with $\Delta m_{K,0}$=0.4 mag. We did not investigate fields with $b>20^\circ $ because we are interested in the disk stars at low $z$; we did not not analyze the brighter stars with $m_{K,0}<9.8$ either, because there are very few to accumulate statistics, nor the fainter stars with $m_{K,0}>13.0$ because they have a higher ratio of dwarf contamination.
We selected only bins with more than ten sources, allowing at most 12\,000 stars per bin (including RCGs and other stars which are not RCGs). In total, the number of RCGs with these criteria is 19\,221\,339 in an area of 14\,100 square degrees, that is, an average of 1\,360 RCGs per square degree.
This sample was also used by L\'opez-Corredoira (2014) to investigate the rotation curve
of the stellar disk.

Then, for each bin of RCGs with a given ($\ell $, $b$, $m_{K,0}$, $(J-K)_0$) we
derived its cumulative extinction along the line of sight and its 3D position in the Galaxy 
($x$, $y$, $z$). The extinction and distance follow (L\'opez-Corredoira et al. 2002b, Sect. 3)
\begin{equation}
A_K=0.67[(J-K)_0-(J-K)_{{\rm red\ clump}}]\ \ {\rm (Marshall\,et\,al.\,2006)}
,\end{equation}
\begin{equation}
r=10^{\frac{m_{K,0}-M_{K,{\rm red\ clump}}+5-A_K}{5}}
.\end{equation}
We used the updated parameter values $M_{K,{\rm red\ clump}}=-1.60$, 
$(J-K)_{{\rm red\ clump}}=0.62$ (averages of the values given by Laney et al. 2012 and
Yaz G\"ok\c ce et al. 2013) and Galactocentric distance of the Sun $R_\odot =8$ kpc (Malkin 2013); 
we neglected the height of the Sun over the plane.
We did not consider any error in the distance determination because of the error in the absolute magnitude
of the RCGs, which is expected to be negligible in the outer disk: in the worst of the cases with $m_K=13.0$,
typical uncertainties of $\lesssim 0.02$ mag in the RCG absolute magnitude 
(Laney et al. 2012; Yaz G\"ok\c ce et al. 2013), plus the determination inaccuracy of the average
absolute magnitude on the order of 0.007 (for a typical number of stars per bin of 
400, and $\Delta m=0.2$), it leads
to relative errors on the distance of 1\% for distances of about 8 kpc. 
For lower magnitudes there are fewer stars per bin, which produces higher relative errors of the distance, but the error produced in the velocities is lower because of the smaller distances. We do not know the extinction uncertainty; assuming errors of up to 0.02 mag in K in the outer disk, we derive the same negligible uncertainties, except perhaps in the very few regions strictly in the plane ($z=0$), where the error might be somewhat larger. In the inner disk, the errors may be larger because of higher extinction, but they are still smaller than other statistical and systematic errors we derive.

\subsection{Contamination}
\label{.contam}

Our selected sources (almost 20 million) are mostly RCGs, but there is also a
fraction of contamination. L\'opez-Corredoira (2014, Sect. 3.1) discussed this
and reached the conclusion that, in the most pessimistic case and
for the faintest stars ($m_K=13.0$), the contamination might reach 20\%, composed of
main-sequence stars and giant stars different from RCGs. For brighter
stars the contamination should be lower.

This contamination is important for calculating the average proper motions, because dwarfs are much closer
than RCGs, and attributing an incorrect distance by a factor $f$ would imply multiplying its
real proper motion by a factor $f$, which in turn would mean 
huge linear velocities for these dwarfs and a considerably increased
average if there is some net average velocity of these dwarfs. 
But this can be improved by using the median of the proper motions instead of
the average of proper motions and by bearing in mind that there might be 
a systematic error due to this contamination, 
which would move the median to the position of the ordered set of proper motions
in the range of 40-60\% instead of 50\%.

\subsection{Proper motion of the RCGs}

From the mean 3D position of the bin with ($\ell $, $b$, $m_{K,0}$) with a number $N$ of RCG stars, we calculated its median proper motion in Galactic coordinates: $\mu _\ell $, $\mu _b$. 
The angular proper motion is, of course, directly converted into a linear velocity
proper motion ($v_\ell $, $v_b$) simply by multiplying $\mu _\ell \cos b$ and $\mu _b$ by its distance from the Sun. Only bins with $N>10$ are included in the calculation. We found the use of the median instead of the average more appropriate because, as said in the previous subsection, it is a better way to exclude the outliers because of all kinds of errors. 
The statistical error bars of the median were calculated for 95\% C.L.: the upper and lower limits 
correspond to the positions $0.5N\pm 0.98\sqrt{N}$ of the ordered set of $N$ data. 
Since the errors of the averages are evaluated by a $\chi ^2$ analysis, the confidence level associated with these error bars is not important at this stage; their inverse
square was just used as weight in the weighted averages of multiple bins, and the error bar of these
averages were quantified from the dispersion.
Moreover, as said above, we calculated a systematic error due to contamination: the upper and lower limits 
correspond to the positions $0.5N\pm 0.1N$ of the ordered set of $N$ data (assuming the worst cases
in which the contamination of non-RCGs is 20\% and that the proper motions of these non-RCGs are all higher
or lower than the median).

\subsection{Correction of systematic errors in the proper motions}
\label{.corrsyst}

Proper motions published by the PPMXL catalog have both statistical and systematic errors.
The transmission of the statistical errors is taken into account in the previous steps; 
however, the systematic errors need to be accounted for as well because they are relatively high.
L\'opez-Corredoira (2014, Sect. 5) calculated these systematic errors of the proper motions ($Syst[\mu _b]$) as a function of the Galactic coordinates using Quasi Stellar Objects (QSOs) as reference of null proper motions and interpolating the value of the systematic error as a function of coordinates.
Then, we subtracted this systematic error from each of our proper motions of our bins:
\begin{equation}
(\mu _b)_{\rm corrected}=\mu _b - Syst[\mu _b](\ell, b)
\label{corrsyst}
.\end{equation}
Note that the error bar of $Syst[\mu _b]$ is also a systematic error, so it cannot be
reduced by increasing the number of sources. In our case, we derive that this error is
still approximately 1 mas/yr in any direction, somewhat lower than the average $|Syst[\mu _b]|$ and,
most importantly, $(\mu _b)_{\rm corrected}$ have an average null deviation with respect to
the true values, whereas $\mu _b$ does not.

\section{Deriving the vertical velocity from proper motions}
\label{.proper}

\begin{figure}
\centering
\includegraphics[width=8cm]{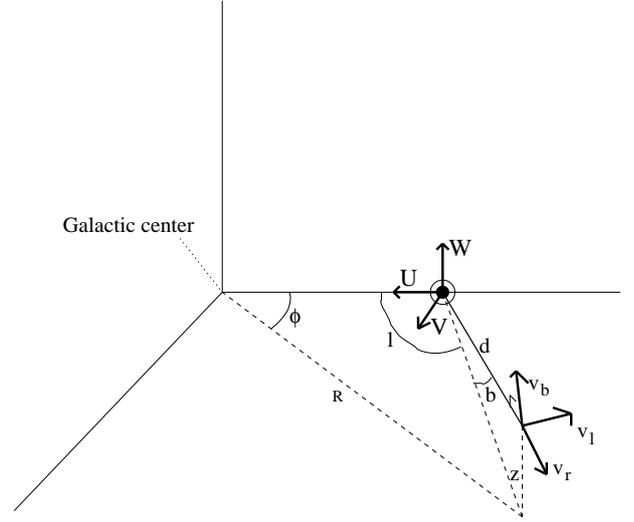}
\caption{Sketch of kinematics of a star with respect to the Sun.}
\vspace{1cm}
\label{Fig:convvel}
\end{figure}

The 3D velocity of the combination of radial velocity ($v_r$) and proper motions ($v_\ell $, $v_b$)
is related to the velocity in the reference system $U$, $V$, $W$ as plotted in Fig. \ref{Fig:convvel}
by
\begin{equation}
v_r=U_*\cos \ell \cos b+V_*\sin \ell \cos b+W_*\sin b
,\end{equation}\[
v_\ell =-U_*\sin \ell+V_*\cos \ell
\]\[
v_b=-U_*\cos \ell \sin b-V_*\sin \ell \sin b+W_*\cos b
,\]
where $(U_*, V_*, W_*)$ is the velocity of a star relative to the Sun in the system (U,V,W).
The Sun velocity in this system with respect to the Galactic center
is $(U_\odot , V_{g,\odot }, W_\odot )$; the second coordinate
\begin{equation}
V_{g,\odot }=\Omega (R_\odot ,z=0)+V_\odot 
,\end{equation}
where $\Omega (R_\odot ,z=0)$ is the rotation speed of the Local 
Standard of Rest (LSR) with respect to the Galactic center;
$(U_\odot , V_\odot , W_\odot )$ is the velocity of the Sun with respect to the LSR.
Here, we adopt the values $U_\odot =14.0\pm 1.5$ km/s, $V_\odot =12\pm 2$ 
km/s and $W_\odot =6\pm 2$ km/s (Sch\"onrich 2012). We used a value of 
$V_{g,\odot }=250\pm 9$ km/s (Sch\"onrich 2012).

We assume that the in-plane velocities of the stars are only those of circular orbits whose rotation speeds vary with the Galactocentric radius $R$. This simplifying assumption can be adopted because we average over large samples, therefore the net velocity of each group of stars is well represented by circular motion. In addition, because we restricted the sample to low Galactic latitudes ($|b|<20^\circ $), the contribution of these in-plane components is further diminished. Hence, we derive the coplanar velocities of the star with respect to the Sun as the projection of the circular speed, $\Omega (R,z)$ (independent of the azimuth $\phi $), along each of the velocity axes,
\begin{equation}
\label{UV}
U_*=-U_\odot +\Omega (R,z)\sin \phi
,\end{equation}\[
V_*=-V_{g,\odot }+\Omega (R,z)\cos \phi
.\]
The effect of the warp in these expressions is negligible (L\'opez-Corredoira 2014,
Sect. 4). 

The vertical velocity of the star in the Galactocentric system is
\begin{equation}
\label{W}
W=W_\odot +W_*
,\end{equation}
where $\phi $ is the Galactocentric azimuth of the star. Hence,
\begin{equation}
W=\frac{v_b}{\cos b}+W_\odot-U _\odot \cos \ell \tan b -V_{g,\odot }\sin \ell \tan b
\label{velvert}
\end{equation}\[
\ \ \ \ \ \ +\Omega (R,z)\sin (\phi +\ell )\tan b 
.\]

Eq. (\ref{velvert}) allows us to determine the vertical velocity only with the determination
of the proper motion in the Galactic latitude projection and the rotation speed. This is what we do here. 
We adopt $\Omega (R,z=0)=238$ km/s $\forall R$. 
Note that, since we are at low $b$, the last term 
in Eq. (\ref{velvert}) containing $\Omega $ is small and consequently $W$ only weakly depends on $\Omega $; therefore small variations of $\Omega $ with respect to the approximation of a flat rotation curve 
do not significantly change the results in $W$.

\subsection{Results}

We are interested in the region $R>4$ kpc, $|z|<2$ kpc, which defines the disk region. For $R<4$ kpc
we find the non axisymmetric structure with non circular orbits of the long bar (Am\^ores et al.
2013). For $|z|>2$ kpc, the halo becomes important and the stellar density of the disk is a factor $\lesssim 300$ lower than at $z=0$ (Bilir et al. 2008). 

\begin{figure}
\vspace{1cm}
\centering
\includegraphics[width=9cm]{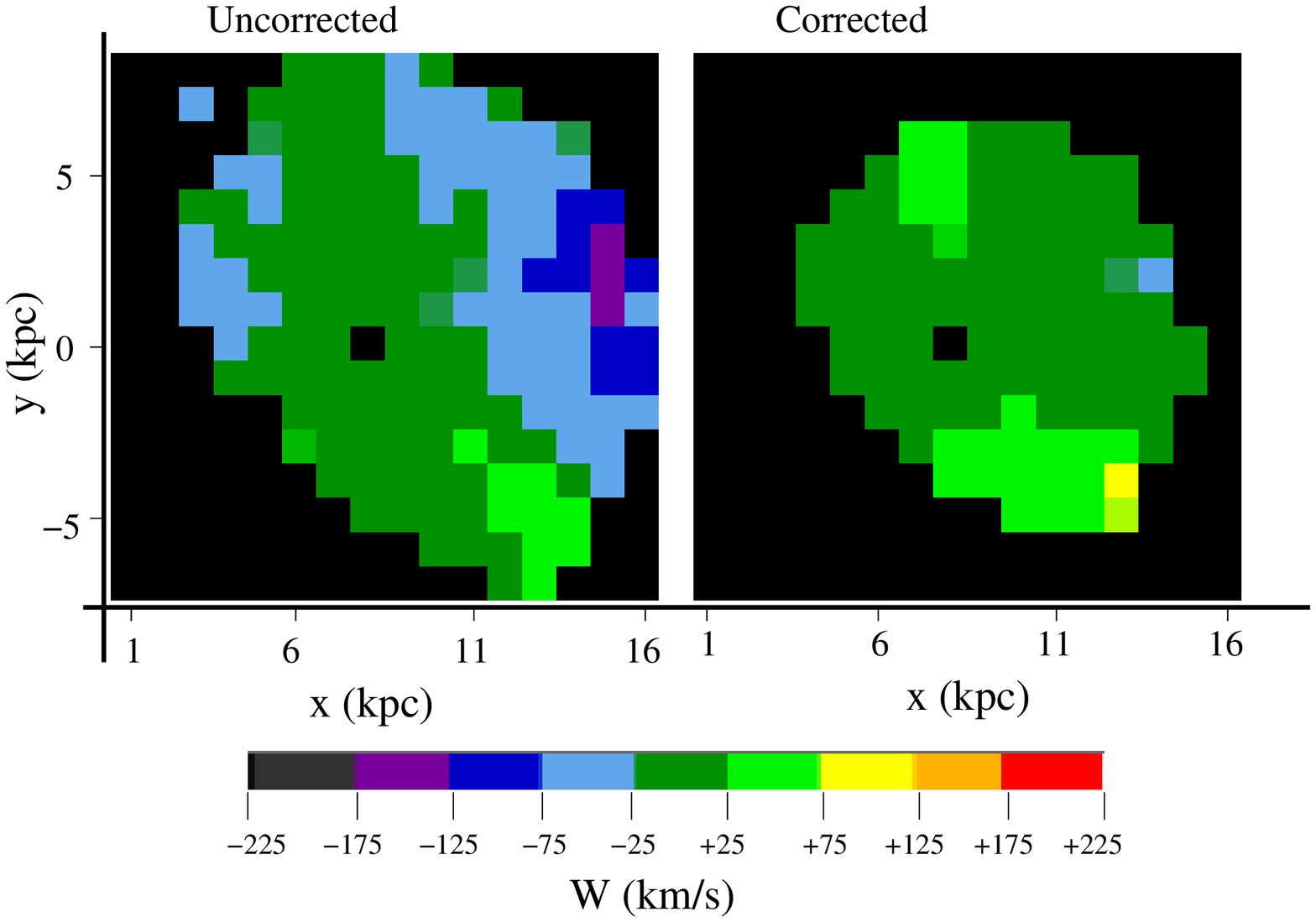}
\caption{Average vertical velocity using PPMXL data 
as a function of Galactocentric Cartesian coordinates $x$, $y$ (the position of the Sun is $x=8$ kpc, $y=0$) for $|z|<2$ kpc. Only bins with error bars lower than 50 km/s are plotted; black indicates larger errors or absence of data. The left panel
is the weighted average of the bins without correcting for systematic errors of the proper motion. The right panel is the weighted average of the bins including the correction for systematic errors 
of the proper motion.}
\vspace{1cm}
\label{Fig:vertical2D}
\end{figure}

The result of the calculation of $W$ according to Eq. (\ref{velvert}) 
is plotted in Fig. \ref{Fig:vertical2D}, including 
a weighted average of all the bins with common $x$, $y$ (Galactocentric
coordinates, with the position of the Sun at $x=8$ kpc, $y=0$).
We plot the velocities $W$ obtained from $\mu _b$ without correcting for systematic errors of the proper motions and with the corrected $(\mu _b)_{\rm corrected}$ from Eq. (\ref{corrsyst}). It is clear that
the correction is substantial and the uncorrected plot has high velocities that are not real.
This was also found in the analysis of the rotation curve by L\'opez-Corredoira (2014), in which
a significant departure, from a local flat rotation curve would be produced if the correction
for systematic errors were not taken into account.
Abedi et al. (2013) have tried to derive the vertical motions from similar
proper motion data also using the red clump stars and derived a conspicuously negative value
of $W$ toward the anticenter instead of the expected positive value; the measurements
of Abedi et al. (2013) probably just reflect the systematic errors of the proper motion and, if they had appropriately
corrected for Eq. (\ref{corrsyst}), they would have obtained results closer to the model predictions. 
The right panel of Fig. \ref{Fig:vertical2D} shows a map of velocities with lower vertical motions
in most of the bins after the correction 
(taking into account that the error bars of each plotted bin are $\le 50$ km/s).
In Figs. \ref{Fig:w_phi} and \ref{Fig:w_z}, we show the results of $W$ 
as a function of $\phi $ within $|z|<2$ kpc or as function of $z$ averaging over all values of $\phi $
for different ranges of $R$.

\begin{figure*}
\vspace{1cm}
\centering
\includegraphics[width=16cm]{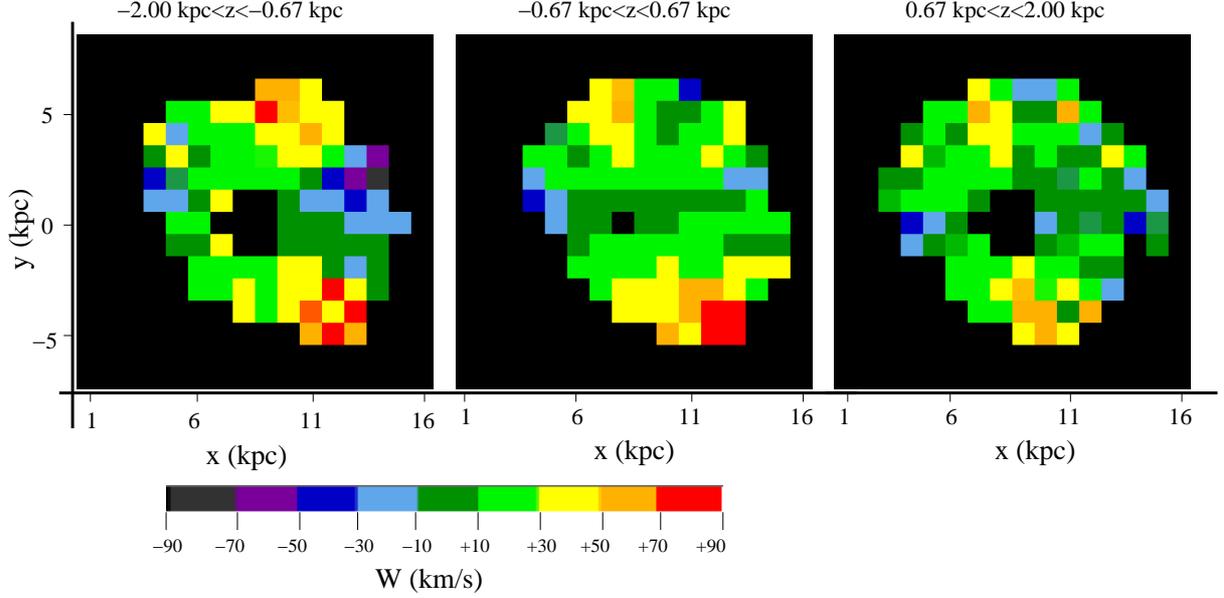}
\caption{Average vertical velocity using PPMXL data, including the correction for systematic errors 
of the proper motion, as a function of Galactocentric Cartesian coordinates $x$, $y$ (the position of the Sun is $x=8$ kpc, $y=0$) for different ranges of $z$. 
Only bins with error bars lower than 50 km/s are plotted; black indicates larger errors or absence of data.}
\vspace{1cm}
\label{Fig:vertical2Dz}
\end{figure*}

\begin{figure*}
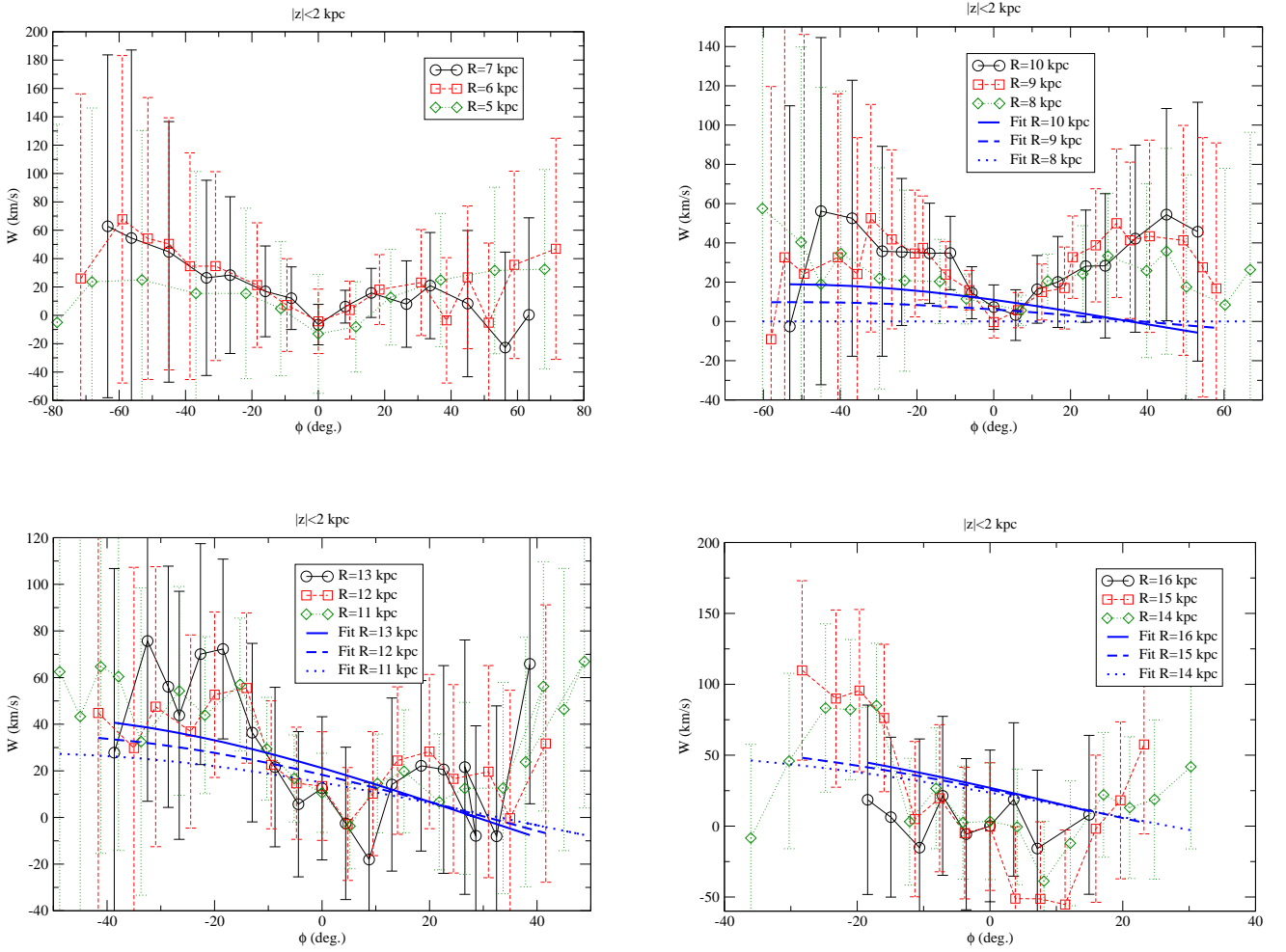

\vspace{1cm}
\centering
\includegraphics[width=8cm]{w_phi5-7.eps}
\hspace{1cm}
\includegraphics[width=8cm]{w_phi8-10.eps}\\
\vspace{1cm}
\includegraphics[width=8cm]{w_phi11-13.eps}
\hspace{1cm}
\includegraphics[width=8cm]{w_phi14-16.eps}\\
\caption{Average vertical velocity as a function of Galactic cylindrical coordinates $R$, $\phi $ for $|z|<2$ kpc. For $R\ge 8$ kpc, we also show the best fit given by Eq. (\protect{\ref{bestfit}}).}
\vspace{1cm}
\label{Fig:w_phi}
\end{figure*}

\begin{figure}
\vspace{1cm}
\centering
\includegraphics[width=8cm]{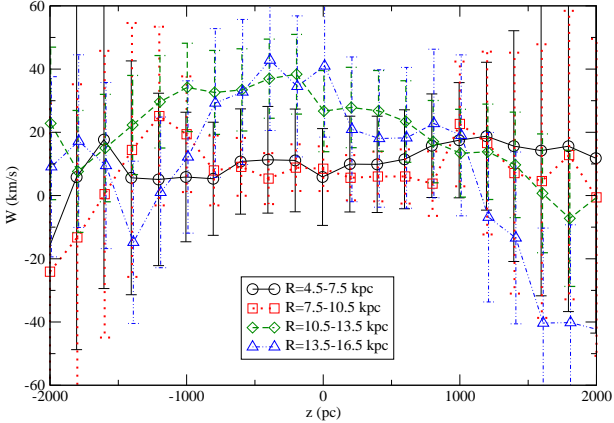}
\caption{Average vertical velocity as a function of the vertical position, $z$, for different ranges
of Galactocentric distances $R$.}
\vspace{1cm}
\label{Fig:w_z}
\end{figure}

In Fig. \ref{Fig:vertical2Dz} we show the same kind of map, but with higher resolution velocity and dividing
the total sample into three subsamples of different $z$. The southern hemisphere map 
($-2<z({\rm kpc})<-0.67$) shows additional deviations from zero velocity.

\begin{figure}
\vspace{1cm}
\centering
\includegraphics[width=8cm]{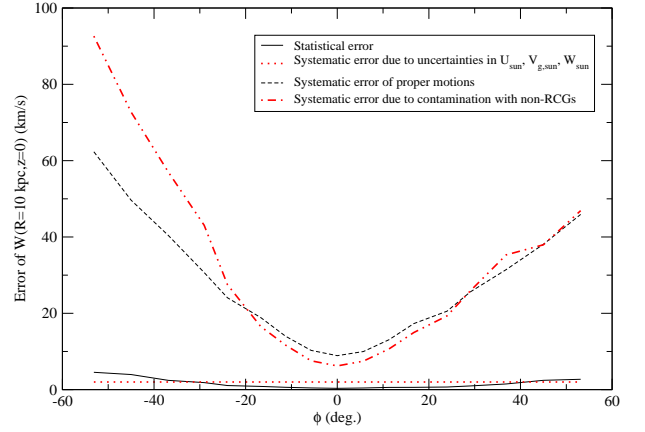}
\caption{Decomposition of the four error sources, which sum quadratically to give
the error bars plotted in Fig. \protect{\ref{Fig:w_phi}} for $R=10$ kpc.}
\vspace{1cm}
\label{Fig:errors_w}
\end{figure}

In these plots the error bars of the different bins are not entirely independent because the systematic errors are not independent. 
The error bars are dominated by the systematic errors because of the contamination of non-RCGs (we have assumed the most pessimistic scenario of a 20\% contamination and that the proper motions of these non-RCGs are all 
higher or lower than the median) and the systematic errors of the proper motions derived using the QSOs reference; see Fig. \ref{Fig:errors_w} for an example of decomposition or errors for $R=10$ kpc. 

\section{Relating the vertical motion to the warp kinematics}
\label{.warp}

If we consider that this vertical motion a result of the warp, modeled as a set of circular rings that are rotated and whose orbit is in a plane with angle $i_w(R)$ with respect to the Galactic plane, then
\begin{equation}
W=\Omega (R,z'=z-z_w)\sin [i_w(R)]\cos (\phi -\phi _w)+\dot{z_w}
,\end{equation}
\begin{equation}
z_w(R,\phi )=R\tan [i_w(R)]\sin (\phi -\phi _w)
,\end{equation}
where $\phi _w$ is the azimuth
of the line of nodes, and $z_w$ is the height of the disk over the $b=0$ plane. We assume
the greatest height of the warp to be 
\begin{equation}
z_w(R>R_\odot,\phi=\phi _w+\pi /2)\approx \gamma (R-R_\odot )^{\alpha }
\label{zwmax}
,\end{equation}
and an invariable line of nodes (extremely slow precession, i.e. $\dot{\phi _w}\ll \dot{\gamma }$) and no change in the shape of the warp. 
We also assume, as above, a constant $\Omega (R,z)=\Omega _{LSR}=238$ km/s; this
may be slightly reduced for high $R$ or high $|z|$ (L\'opez-Corredoira 2014), but the
order of magnitude does not change, so $W$ is only weakly affected by a change
of the rotation speed. 
Joining these assumptions, we derive, for low angles $i_w(R)$,
\begin{equation}
W(R>R_\odot , \phi ,z=0)
\approx \frac{(R-R_\odot )^\alpha}{R}[\gamma \omega_{LSR}\cos (\phi -\phi _w)
\end{equation}\[
\hspace{3cm} +\dot{\gamma }R\sin (\phi -\phi _w)]
.\]

We adopt the values $\alpha =1$ (Reyl\'e et al. 2009), $\phi _w=+5\pm 10$ deg, in the middle of the range
of azimuths given in the literature for the stellar warp between -5 and +15 deg. 
(L\'opez-Corredoira et al. 2002b; L\'opez-Corredoira 2006; Momany et al. 2006; Reyl\'e et al. 2009), 
and we use our regions with $R\ge 8$ kpc to obtain the best fit:
\begin{equation}
\label{bestfit}
W_{\rm best\,fit}=(54\pm 38\,{\rm km/s})\left(1-\frac{R_\odot }{R}\right)[\cos (\phi -\phi _w)
\end{equation}\[
\hspace{3cm} -(0.14\pm 0.07\,{\rm kpc}^{-1})R\sin (\phi -\phi _w)]
.\]
$\chi ^2=68.5$ for $N=153$.
The errors include the error derived from a $\chi ^2$ statistical analysis (for 68.3\% C.L.)
and the uncertainty in $\phi _w$. The function is plotted in Fig. \ref{Fig:w_phi}.
Hence, the values of $\gamma $ and $\dot{\gamma }/
\gamma $ that fit our data are
\begin{equation}
\label{gamma}
\gamma =0.23\pm 0.16
,\end{equation}
\begin{equation}
\frac{\dot{\gamma }}{\gamma }=-34\pm 17\,{\rm Gyr}^{-1}
\label{gammadot}
.\end{equation}

If we assume an exponent $\alpha =2$ instead of $\alpha =1$ we get:
$\gamma =0.032\pm 0.024$ kpc$^{-1}$, $\frac{\dot{\gamma }}{\gamma }=-49\pm 29$ Gyr$^{-1}$,
which shows the same trend of decreasing amplitude of the warp with similar
frequency. The observed trend only mildly depends on the assumed shape of the warp. 
The lower $\chi ^2$ for $\alpha =2$ is 84.2 ($N=153$), higher than for $\alpha =1$, so 
we assume the $\alpha =1$ values in the rest of this paper.

These results (for $\alpha =1$) can be interpreted as follows:
\begin{enumerate}
\item Our data are not good enough to trace the structure of the stellar warp. There are much
better methods to derive the morphology with positions and velocities of the warp stars, for example, that of Abedi et al. (2014). At least we are able to derive a range of $\gamma $ that is compatible with other measurements of the stellar warp height. For instance, from Eq. (\ref{zwmax}), we compute a maximum height of $0.92\pm 0.56$ kpc at $R=12$ kpc or $1.61\pm 0.98$ kpc at $R=15$ kpc, compatible with the values obtained by L\'opez-Corredoira et al. (2002b), Momany et al. (2006), or Reyl\'e et al. (2009).

\item We can derive some information of the warp kinematics in adition to
the circular motions of the stars around the Galactic center. Eq. (\ref{gammadot}) indicates
that our warp is not stationary ($\dot{\gamma }=0$), although only at 2$\sigma $.
If we assume a sinusoidal oscillation, $\gamma (t)=\gamma _{\rm max}\sin (\omega t)$, we have
a period
\begin{equation}
T=\frac{2\pi }{\omega }=2\pi \left(\frac{\dot{\gamma }}{\gamma }\right)^{-1}\cot (\omega t)
\label{t}
,\end{equation}
and the probability of having a period $T$ is the convolution of two probability distributions:
the Gaussian probability as a result of the error bar of $\dot{\gamma}/\gamma $; and the probability proportional
to the amount of time $\Delta t$ in which we can observe values corresponding 
to $\dot{\gamma}/\gamma $ and a period between $T$ and $T+\Delta T$, which
is proportional to $\left|\frac{d(\omega t)}{dT}\right|\left(\frac{\dot {\gamma}}{\gamma }={\rm constant}\right)=\frac{1}{2\pi }\left|\frac{\dot {\gamma}}{\gamma }\right|\frac{1}
{1+\left(\frac{T\dot {\gamma}}{2\pi \gamma}\right)^2}$. 
The normalized convolution of these two probabilities gives 
\begin{equation}
P(T)dT=\frac{dT}{2^{1/2}\pi^{5/2}\sigma _x}\int _{-\infty}^{+\infty} dx \frac{|x|}{1+\left(\frac{Tx}{2\pi }
\right)^2}e^{-\frac{(x-x_0)^2}{2\sigma _x^2}}
,\end{equation}
where $x_0\equiv \frac{\dot {\gamma}}{\gamma}$ and $\sigma _x$ is its r.m.s.
Fig. \ref{Fig:probt} shows this probability distribution. 

From this distribution, 
the cumulative probabilities of 0.159, 0.500, and 0.841 are given for 
$T=0.047$, 0.208, and 1.049 Gyr, respectively, so we can say that $T=0.21^{+0.84}_{-0.16}$ Gyr (68.3\% C.L.),
or $T=0.21^{+10.56}_{-0.20}$ Gyr (95.4\% C.L.). Alternatively, we can say
that $T<0.43$ Gyr (68.3\% C.L.), $T<4.64$ Gyr (95.4\% C.L.).

Our results are equivalent to a rotation of the rings around the line of nodes with
an angular velocity of 
$\frac{\dot {z_w}(R>R_\odot,\phi=\phi _w+\pi /2)}{R}\sim 
-8\left(1-\frac{R_\odot }{R}\right)$ km/s/kpc, which is
between -1 and -4 km/s/kpc in the range of $R$ between 9 and 16 kpc.
The negative value means a decreasing amplitude of the warp with time.
This result disagrees with the result of +4 km/s/kpc given by 
Miyamoto et al. (1993) or Miyamoto \& Zhu (1998) at short distances from
the Sun, but it is more similar to the result of 
-4 km/s/kpc by Bobylev (2010), who also used red clump giants.
Bobylev (2013) obtained with Cepheids a value of -15 km/s/kpc, which
is different from our result. We do not know the reason for this last disagreement;
indeed, our plot of Fig. \ref{Fig:w_phi} for $R=8-10$ kpc is similar to the plot of
$W(Y)$ in Fig. 3 of Bobylev (2013). If Bobylev (2013) had interpreted
his term of $\frac{\partial V}{\partial z}$ in a different way, in terms
of a variation of the rotation speed with $z$, for instance (L\'opez-Corredoira 2014), and
so considering null the deformation tensor, this would lead to an angular
speed of -2 km/s/kpc, compatible with what we see.

\begin{figure}
\centering
\vspace{1cm}
\includegraphics[width=7.5cm]{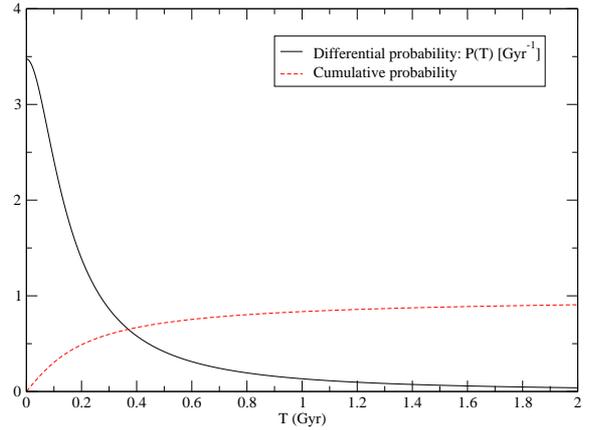}
\caption{Distribution of probability of the period for the motion of $\gamma (t)=
\gamma _{\rm max}\sin (\omega t)$ given that we have 
observed $\frac{\dot{\gamma }}{\gamma }=-34\pm 17\,{\rm Gyr}^{-1}$.}
\vspace{1cm}
\label{Fig:probt}
\end{figure}

\end{enumerate} 

\subsection{Toward a more realistic warp kinematic model}
\label{.model}

Abedi et al. (2014) introduced a new kinematic model 
for the Milky Way warp. Using a test particle simulation,  
a realistic Galactic potential, and a geometrical model for the warp,
the disk potential was calculated and stars were let to evolve in it. This way, a warped
 sample of stars was generated  that not only follow the warped disk, but their kinematical information also reflects the 
signature of the warp. In this model the particles evolve in an adiabatic regime, that is, the warp grows slowly enough so that the stars remain in statistical equilibrium with the potential (hereafter model A).

The PPMXL proper motions (see Sect. \ref{.warp}) 
suggest that the Galactic warp is probably not in a steady state and
 its amplitude can change rapidly with time. From equations (\ref{gamma}) and (\ref{gammadot}), 
we obtain a negative value for $\dot{\gamma}$, which implies 
that the warp amplitude decreases with time. In this subsection, we present a first attempt to analyze the PPMXL data in the context of this new kinematic warp model. We simulated a warped sample of RCGs (see Sect. 3 of Abedi et al. 2014). We first grew the warp in the Galactic disk adiabatically during a time $t_1 = 3.5$ Gyr until it reached a tilt angle of $\psi_1= 16^\circ$ at the Galactocentric radius of 16 kpc, where
$\psi \equiv \frac{z}{R}$. Then, we decreased its amplitude impulsively, following Eq. 3 of Abedi et al. 2014, during a time $t_2=100$ Myr to achieve a tilt angle of  $\psi_2= 6^\circ$ at the Galactocentric radius of 16 kpc (hereafter model B), to quantitatively 
mimic the rapid warp evolution obtained in section 5. We set the tilt angle to be a
 linear function of the Galactocentric radius ($\alpha=1$ in Eq. 1 of Abedi et al. 2014).
We performed this simulation for $56\times 10^6$ RCGs. This is the total number of RCGs in the
 Galactic disk according to the Besancon Galaxy Model (Czekaj et al. 2014). 
We defined the line of nodes to coincide with
 the Sun-Galactic center line ($\phi_w =0$) in the simulations.

\begin{figure}[hp] 
\centering
\includegraphics[scale=.45]{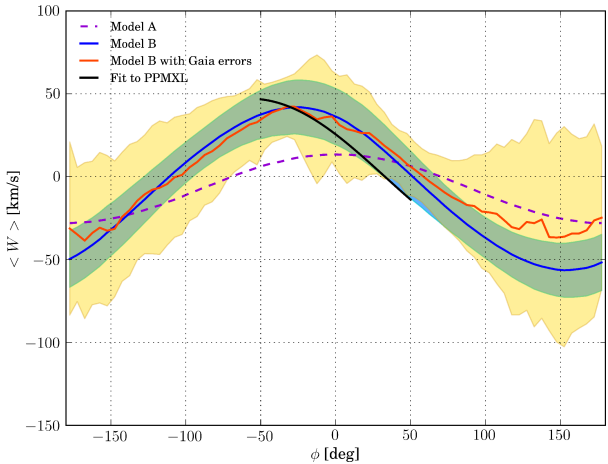}
\caption{Mean W velocity component as a function of galactocentric azimuth for RC stars at $13<R$(kpc)$<14$. RCGs simulated using models A and B are plotted as dashed purple and solid blue lines. The fit to PPMXL data is given in black, the same as the fit in Fig. \ref{Fig:w_phi} in the bottom-left panel. The Gaia 'observed' values are plotted in red. The shaded regions in blue and yellow represents the standard deviations of the $W$ velocity for model B without and with Gaia errors. The line of nodes is defined to coincide with the Sun-Galactic center line 
($\phi_w=0$).}
\label{fig:2}
\end{figure}

In Fig. \ref{fig:2} we plot the resulting $W$ velocity component as a function of Galactocentric azimuth for RCGs at $R$ between 13 and 14 kpc. To show the global trend, we plot the $W$ for the whole 
range in azimuth, whereas in Fig. \ref{Fig:w_phi} we just plotted the range for which we have enough data. To facilitate the comparison,
 we only show this plot for one of the radius bins presented in Fig. \ref{Fig:w_phi}. As expected, for a simulation where the warping occurs in an adiabatic regime, that is, 
stars remain in 
statistical equilibrium with the warped potential (model A), the highest peak of $W$  is always observed
 in the direction of the line of nodes. Whereas for model B, this peak moves towards negative Galactic azimuths and
 gains a larger amplitude. We checked that its amplitude and azimuthal position depends on $t_2$ and $\psi_2$; increasing them
 causes the amplitude to become larger and it moves toward the more positive azimuths. We also have checked that a deviation of $5^\circ$ of the line of nodes from the Sun-Galactic center line will cause the strongest peak seen in Fig. \ref{fig:2} to be slightly shifted by
$\sim 5^\circ$ toward the positive azimuths, which is negligible for our qualitative study. The blue-shaded regions show the errorbars in $W$ for model B that are standard deviations, representing the intrinsic velocity dispersion after evolution in the warped potential.
The black line shows the fit to PPMXL data, the same as the fit seen in Fig. \ref{Fig:w_phi}in the bottom-left panel. The simulations qualitatively follow the same trend as the fit. We checked that by increasing $t_2$ by a few more hundred 
million years, the amplitude of the peak will decrease and it will move toward the more positive azimuths.
Therefore, this very impulsive change in the amplitude of the warp ($t_2=100$ Myr) is a very important parameter for obtaining the same trend as we found for the PPMXL fit.
With higher radii for the simulation of model B, the amplitude of  
the peak in $W$ increases until it reaches $\sim 60$ km/s at $R=16$ kpc,
which is very similar to the result from the PPMXL fit for this radius (see Fig. \ref{Fig:w_phi} bottom right panel).

In a perturbation on the Galactic disk in such a short dynamical time-scale, the stars will not remain in equilibrium with the imposed warp and will cease to show the 
behavior seen in Fig. \ref{fig:2} after evolving a few more orbital periods.
This clearly indicates that this perturbation must be a transient phenomenon.

\section{Relating the vertical motion to vertical waves in the Galactic disk}
\label{.waves}

Widrow et al. (2012) reported some oscillating fluctuations in the
density of stars and vertical velocity $W$ versus the vertical position $z$, and hence
they suggested that there might be vertical waves in the Galactic disk that might have
been produced by the passage of a satellite galaxy or dark matter subhalo through
the Galactic disk; or even by an outer vertical resonance (Griv et al. 2002). The vertical resonance means that resonant coupling between natural oscillations perpendicular to the equatorial plane of the Galaxy and periodic changes in the force toward a galactic center can produce significant out-of-plane 
motions of stars, in which the variation of the force toward the Galactic center may be caused 
by the non axisymmetric potential. The observations of Widrow et al. are doubtful: first because they
used photometric parallaxes to calculate distances, which is subject to many possible
errors that might lead to some small spurious fluctuation in the density 
like those observed in their Fig. 2; and second because in their Fig. 3 no clear
oscillation of $W$ beyond the error bars is observed. Nonetheless,
vertical waves are a possible explanation to account for
certain values of $W$ different from zero.

In Fig. \ref{Fig:w_z} we do not see any oscillation, which would be expected if the
vertical waves existed, although our error bars are quite high to exclude it. Only in the
range $R=7.5-10.5$ kpc do we obtain relatively small error bars but even then we do not see
oscillations. We only see values of $W$ between -20 and +20 km/s without any trend, 
similar to those obtained by Widrow et al. (2012) or Williams et al. (2013).  

For $R>10.5$ kpc in Fig. \ref{Fig:w_z} we see a north-south asymmetry:
some values of $W$ lie significantly higher than zero for $-1\lesssim z({\rm kpc})\lesssim 0$. This might indicate that there is some vertical wave. However, because we do not observe
this at lower $R$, this possible oscillation most likely has something
to do with the warp, and in particular with the southern warp (with $z<0$, $\phi <0$). 
In Figs. \ref{Fig:w_phi} and \ref{Fig:w_z} we observe that the significant
departure of $W=0$ is precisely for $z<0$, $\phi <0$. 
Fig. \ref{Fig:vertical2D} shows at the location of the southern warp is: at $\ell \approx 240^\circ $. This is reinforced in Fig. \ref{Fig:vertical2Dz} for $z$ between -2 and -0.67 kpc, where the southern warp deviates more strongly from zero vertical velocity than in the northern
map. Indeed, we know the Galactic warp presents a distortion in the southern part in the form of an S-shaped warp (Levine et al. 2006): for $R\gtrsim 15$ kpc, the vertical shift of the southern warp becomes constant or decreases whereas the northern warp shift continues to grow; it is expected that this anomaly is 
produced by some transient perturbation, whereas the S-shaped warp may well be a long-lived feature
($\dot{\gamma }=0$), which agrees with the fact that Figs. \ref{Fig:w_phi} and \ref{Fig:w_z} show no strong deviations from $W=0$. 
The Canis Major overdensity is also placed in this position, but it was clear from the
discussions in previous papers (Momany et al. 2006; L\'opez-Corredoira 2006; L\'opez-Corredoira et al.
2007) that this overdensity is precisely a result of the southern warp and not of any extragalactic element or any other substructure within our Galaxy.

\section{Improving the accuracy of our results}
\label{.improve}

The accuracy of our results is low, and we only report a tentative detection of
vertical motion with low significance. One may wonder how this result might be improved.
From Fig. \ref{Fig:errors_w}, we understand the origin of our large error bars: the statistical errors are low (we have 20 million RCGs, so this is expected), and the
inaccuracy of the solar motions with respect to the LSR is also unimportant; but there are
two strong error sources: 1) the systematic errors of the proper motions, even after
correcting them with the quasars reference; and 2) the contamination of non-RCGs.
The accuracy of the systematic errors in the proper motions will be much improved in future
surveys in the visible, mainly with Gaia (Perryman et al. 2001, Lindegren et al. 2008), 
which will also have many millions of RCGs among their sources. 
Hence, we must only wait for these data to be able to significantly decrease the
systematic error of proper motions. To avoid the contamination
of non-RCGs spectroscopic data (Bovy et al. 2014) or asteroseismology (Kallinger et al. 2010) might be used.
In the following subsections, we analyze a subsample of RCGs selected spectroscopically within PPMXL
and the expectations of applying this method either with photometric or spectroscopic
selections of RCGs for future data with Gaia.

\subsection{Calculations with RCGs selected spectroscopically with APOGEE data}
\label{.apogee}

We repeated our calculations with RCGs selected spectroscopically with APOGEE data. 
The APOGEE is an H-band high-resolution spectroscopic survey (Eisenstein et al. 2011) 
of the third stage of SDSS project (Gunn et al. 2006) (SDSS-III). A detailed description of the target selection and data reduction pipeline is presented in Zasowski et al. (2013). For the purposes of this paper, we used the spectroscopically selected sample of red clump stars presented by Bovy et al (2014).

The subsample of APOGEE contains 10\,341 RCGs (originally 10\,352, but eleven of them were duplications), 10\,297 of which have PPMXL proper motions and 9\,184 of which
are within $|b|<22.5^\circ $. With the same analysis as above, but in bins of $\Delta \ell \times \Delta b=5\,{\rm deg}\times 5\,{\rm deg}$, we obtained the results of Figs. \ref{Fig:w_phiAPO}.
Fig. \ref{Fig:errors_wAPO} shows an example of error decomposition 
for $R=10$ kpc: in this case the error due
to contamination is absent, but instead a large bar is associated with the statistical error, because there are only nine thousand sources instead of 20 million in the total PPMXL. 
Because of this still very large error
bar and also because we are constrained within shorted distances, so that we have no bins (with more than 10 RCGs) for $R>14$ kpc, our results of the warp kinematics are poorer than with the whole PPMXL sample. In particular, with this spectroscopic RCGs sample, we obtain
\begin{equation}
\gamma =0.29\pm 0.17
,\end{equation}
\begin{equation}
\frac{\dot{\gamma }}{\gamma }=-10\pm 35\,{\rm Gyr}^{-1}
,\end{equation}
$\chi ^2=19.1$ for $N=37$; 
which is compatible with the previous results, but with a much higher error bar for $\dot{\gamma }/\gamma $. The conclusion then is that it certainly would be ideal to have a spectroscopic determination
of the red clumps classification to avoid contamination, but to be competitive we would need 1) many more RCGs, at least one order of magnitude more (around 100\,000 RCGs or more) and 2) a
high ratio of these RCGs at $R>14$ kpc, which means that we need to reach fainter magnitudes than APOGEE.

\begin{figure*}
\vspace{1cm}
\centering
\includegraphics[width=8cm]{w_phi5-7APO.eps}
\hspace{1cm}
\includegraphics[width=8cm]{w_phi8-10APO.eps}\\
\vspace{1cm}
\includegraphics[width=8cm]{w_phi11-13APO.eps}
\hspace{1cm}
\includegraphics[width=8cm]{w_phi14-16APO.eps}\\
\caption{Same as Fig. \protect{\ref{Fig:w_phi}} but for the subsample of 9\,186 RCGs selected spectroscopically with APOGEE data.}
\vspace{1cm}
\label{Fig:w_phiAPO}
\end{figure*}


\begin{figure}
\vspace{1cm}
\centering
\includegraphics[width=8cm]{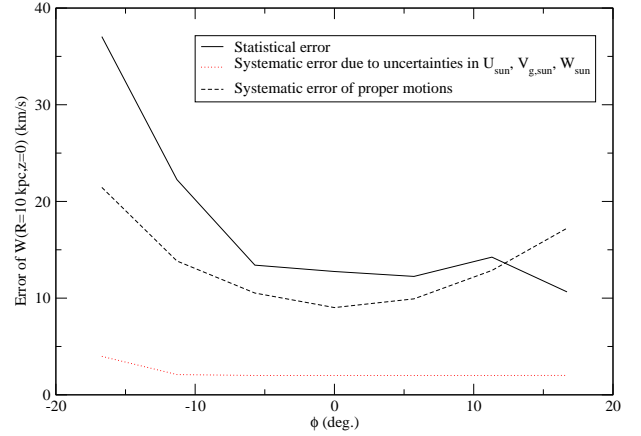}
\caption{Decomposition of the three error sources, which sum quadratically to give
the error bars plotted in Fig. \protect{\ref{Fig:w_phiAPO}} for $R=10$ kpc.}
\vspace{1cm}
\label{Fig:errors_wAPO}
\end{figure}

\subsection{Estimating the accuracy with future Gaia data}

The expected results of the Gaia mission (Perryman et al. 2001, Lindegren et al. 2008) are the most promising candidates to help resolving the ambiguities in the (vertical) motion of the stars in the Galaxy. Gaia, a cornerstone ESA satellite, will produce a 3D map of the stellar content of the Galaxy by mapping about 1 billion objects of the Galaxy and beyond with unprecedent combination of sensitivity and spatial and spectral resolution, thus producing a catalog that will contain around 1\% of the Galactic stellar sources.

Expected errors from the combined used of the Gaia Universe model and Gaia object generator, both part of the Gaia simulator, have recently been calculated and compiled by Luri et al. (2014). In that paper, the errors in the different parameters measured by Gaia are given as fractions of the parallax error. These errors include the random effect of the measurement system on board of Gaia, as they were modeled and introduced in the computation. In addition, the hypothesis that the systematic errors in the measurements are in lower magnitude than the statistical errors is assumed throughout the paper. These systematic errors are introduced by several instrumental artifacts that affect the precision of the measurements: optical aberrations, non-random variations of the detector sensitivity, poor or unstable wavelength calibration, and so on. The systematics do not decay with the use of large target sample as do the statistical errors, and therefore they are the main contributor to the final error in the average velocities. Moreover, the apparent motion of quasars, considered as a reflection of the observer’s motion with respect to the distant Universe, has been simulated in the Gaia Universe model (Robin et al 2012), and so is included in the reported results of Luri et al. (2014).

Following Luri et al (2014, Table 2), the mean value of proper motion errors at the end of the mission ranges from 70 to 80 $\mu $as/yr, considering both $\mu_{\alpha}\cos \delta $ and $\mu_{\delta}$. The mean error of the radial velocity at the end of mission will be 8 km/s. While these mean errors should not be used directly because the computed error distribution is not symmetrical, they can serve as a conservative estimate of the systematic error of the result of the Gaia mission. This estimate is even more conservative if one takes into account that the median G magnitude of the sample in Luri et al. (2014) is 18.9 mag. and that the Gaia error in proper motion shows a remarkable magnitude equation (Luri et al. 2014, Fig. 2). Assuming an approximate equivalence between the G and V magnitudes and an intrinsic (V-K) color for the red clump stars of $<2.5$ mag (Alves 2000), the cut in K-magnitude of our sample will place it safely 1 or 2 magnitudes brighter than the stated value for the G median magnitude, hence decreasing the mean error value in proper motion well below the mean values of Luri et al. (2014).

Under this assumption, the Gaia results will produce a gain in precision in the vertical velocities of a factor of 12-15, because the systematic errors will decay from about 1 mas/yr in the PPMXL (after applying the correction of Sect. \ref{.corrsyst})
to the quoted 70 to 80 $\mu $as/yr in Gaia. Thus the contribution of the systematics in Fig. \ref{Fig:errors_w}, for example, will drop to a level of a few km/s. This will yield an improvement of about $\sqrt{2}$ in the final error, which will be largely dominated by the contamination caused by the poor identification of red clump sources.

If, in turn, we combine a catalog with clearly identified sources as the spectroscopically selected red clumps in APOGEE (see Sect. \ref{.apogee}), then the errors in the vertical velocities can be strongly decreased, even to the level of the systematic error of the Gaia measurement discussed above. In spite of the paucity of sources, the precision of the vertical velocity estimate can be restricted to a few km/s, but the effect of the lack of spatial coverage in determining the overall shape of the vertical velocity of the full warp structure
still remains.

Moreover, the trigonometric parallaxes provided by Gaia will help us improve the accuracy in distance estimation. Using the nominal Gaia error models, we have calculated that toward the Galactic anticenter a relative error in parallax of better than 10\% can be reached at a Galactocentric distance of about 13 kpc. This calculation was made considering the interstellar extinction (Drimmel et al. 2003). However, for stars with larger trigonometric parallax errors, a combination of photometric distances and trigonometric parallaxes will enhance the distance estimate. In addition, knowing more accurate distances will decrease the non-RCGs contamination in the sample. For instance, dwarf contamination can easily be removed because they are located at much smaller distances than RCGs.

To determine the observed trend of the warp kinematic model used in Sect.
\ref{.model} when Gaia data are used, we applied the Gaia observational constraints to the simulated model B sample. Using the nominal Gaia performances, we calculated the errors in trigonometric parallax and proper motions (see Abedi et al.
 2014 for details). The 'observed' values in Fig. \ref{fig:2} can follow the true values except 
for the regions toward the far side of the Galaxy where the interstellar extinction is very high and accordingly, there are fewer stars 
per longitude bin. The position of the peak in $W$ can be clearly detected in 
the observed values.

\section{Conclusions}
\label{.concl}

Our analysis of the proper motions with the PPMXL survey shows a tentative detection [within $\sim 2\sigma$, according to the result of Eq. (\ref{gammadot})] of the vertical oscillation produced in the
southern warp, which tends to decrease its amplitude ($\dot {\gamma }<0$, i.e., $W>0$ for $z_w<0$).
A simple model with $z_w(R,\phi )=\gamma (R-R_\odot )\sin (\phi -\phi _w)$ is well fitted with 
$\gamma =0.23\pm 0.16$, $\frac{\dot{\gamma }}{\gamma }=-34\pm 17\,{\rm Gyr}^{-1}$.
There are two contributions of the warp to the vertical motion: one produced by the inclination of the orbits, another contribution from the variation of $\gamma $. We were able to detect, both analytically (Sect. \ref{.warp} and with simulations (Sect. \ref{.model}),
that the second factor is necessary to fit our data.
If we assume this detection of $\dot{\gamma }\ne 0$ to be real, the period of this oscillation is shorter than 0.43 Gyr at 68.3\% C.L. and shorter than 4.64 Gyr at 95.4\% C.L., excluding at a high confidence level the slow variations ($T>5$ Gyr) that correspond to long-lived features. 
But because we observe this vertical motion only in the southern warp, this most likely indicates that the main S-shaped structure of the warp is a long-lived feature, whereas the perturbation that produces an irregularity in the southern part is most likely a transient phenomenon. 
Moreover, the simulations used in Sect. \ref{.model} indicate that the Galactic disk potential is perturbed on a very short time scale, that is, in an impulsive regime, and therefore stars cannot remain in statistical equilibrium with the potential. This will cause the observed southern warp signature to be lost in a few orbital periods. 

Higher precision data, which surely will be available in the near future, are necessary to resolve this problem. With the accuracy of the present-day data one cannot go proceed.
In future analyses with upcoming surveys such as Gaia together with spectroscopic follow-up of some sources such as
the present-day APOGEE or Gaia-ESO survey (Randich et al. 2013), a much higher accuracy is expected for these results. Our present measurements of $W$ in the farthest part of the disk are poor, with errors of several tens of km/s, but the errors can be reduced by an order of magnitude if we use the future Gaia data plus a spectroscopic classification of red clump giants up to distances of $R=16$ kpc. This will surely reveal the true character of the vertical motion measured here.

\begin{acknowledgements}
Thanks are given to Luis A. Aguilar (Inst. Astronom\'\i a, UNAM, M\'exico) for helpful comments on the simulations of our Sect. \ref{.model}. Thanks are given to the anonymous referee for 
helpful comments and to Astrid Peter (language editor of A\&A) for proof-reading of the text.
MLC and FGL were supported by the grant AYA2012-33211 of the Spanish Ministry of 
Economy and Competitiveness (MINECO).
MLC acknowledges the support of the Fundaci\'o Bosch i Gimpera of Universitat de Barcelona, which 
supported the travel costs of his visit to Univ. de Barcelona.
HA and FF acknowledge the Gaia Research for European Astronomy Training (GREAT-
ITN) network funding from the European Union Seventh Framework Programme,
under grant agreement n$\degr$264895,
and the MINECO (Spanish Ministry of Science and Economy) - FEDER
through grants  AYA2012-39551-C02-01 and CONSOLIDER CSD2007- 00050,
ESP2013-48318-C2-1-R and CONSOLIDER CSD2007-00050. Simulations were carried out
using ATAI, a high performance cluster, at IA-UNAM.
This work has used the data of PPMXL catalog (Roeser et al. 2010) and
the SDSS-III/APOGEE red clump catalog (Bovy et al. 2014). SDSS-III is managed by the Astrophysical Research Consortium for the Participating Institutions of the SDSS-III Collaboration including the University of Arizona, the Brazilian Participation Group, Brookhaven National Laboratory, Carnegie Mellon University, University of Florida, the French Participation Group, the German Participation Group, Harvard University, the Instituto de Astrofisica de Canarias, the Michigan State/Notre Dame/JINA Participation Group, Johns Hopkins University, Lawrence Berkeley National Labora- tory, Max Planck Institute for Astrophysics, Max Planck Institute for Extraterrestrial Physics, New Mexico State University, New York University, Ohio State University, Pennsylvania State University, University of Portsmouth, Princeton University, the Spanish Participation Group, University of Tokyo, University of Utah, Vanderbilt Uni- versity, University of Virginia, University of Washington, and Yale University.
\end{acknowledgements}

\end{document}